# Topological one-way large-area waveguide states in magnetic photonic crystals

Mudi Wang[1], Ruo-Yang Zhang[1], Lei Zhang[1,2,3], Dongyang Wang[1], Qinghua Guo[1,4], Zhao-Qing Zhang[1], and C. T. Chan[1*]

[1]Department of Physics, The Hong Kong University of Science and Technology, Hong Kong, China

[2]State Key Laboratory of Quantum Optics and Quantum Optics Devices, Institute of Laser Spectroscopy, Shanxi University, Taiyuan 030006, China

[3]Collaborative Innovation Center of Extreme Optics, Shanxi University, Taiyuan 030006, China

[4]Institute for Advanced Study, The Hong Kong University of Science and Technology, Hong Kong, China

We have theoretically and experimentally achieved large-area one-way transport by using heterostructures consisting of a domain of an ordinary photonic crystal (PC) sandwiched between two domains of magnetic PCs. The non-magnetized domain carries two orthogonal one-way waveguide states which have amplitude uniformly distributed over a large-area. These two waveguide states support unidirectional transport even though the medium of propagation is not magnetized. We show both experimentally and numerically that such one-way waveguide states can be utilized to abruptly narrow the beam width of an extended state to concentrate energy. Such extended waveguide modes are robust to different kinds of defects, such as voids and PEC barriers. They are also immune to the Anderson type localization when large randomness is introduced.

*Correspondence to: phchan@ust.hk

One-way transport in Quantum Hall systems [1-6], originating from time-reversal symmetry (TRS) breaking, is one of the most striking phenomena in condensed-matter physics. Their counterparts in classical wave systems have also been widely studied both in theory [7-9] and experiment [10-16]. However, topological principles can only guarantee the existence of a certain number of transport channels [17,18]. How well or how fast (group velocity) those channels can transport information or energy depends on the structural details. It is highly desirable if the transport characteristics of topological modes can be controlled.

For a photonic crystal (PC) carrying Dirac points, electromagnetic (EM) wave can propagate in any directions in the bulk. Here, we show that when a topologically trivial PC is sandwiched [19,20] between two topological crystals in which TRS is broken by two oppositely biased magnetic fields, the topologically trivial PC can exhibit unidirectional transmission that is robust to imperfections and defects. This type of heterostructure permits one-way transport of waves in a large area, which has some advantages over topological edge modes [7-16] where the wave energy is confined in a narrow boundary. For example, topological one-way beam spatial collimation can be easily achieved and we have a large area to collect incident wave energy and many degrees of freedom to control wave impedance.

The heterostructure consists of three domains, labelled as A, B and C by three different colors, as shown in Fig. 1(a). Each domain is a PC made up of a hexagonal array (16 mm lattice constant) of cylindrical yttrium iron garnet (YIG) rods (4 mm diameter, 10 mm height). The YIG rods are magnetized by external magnetic field along positive (negative) vertical direction in domain A (C), and no magnetic field is imposed in domain B. The detailed experimental setup is given in the Supplementary Material [21]. The band structures of each domain, calculated using COMSOL, are shown in Fig. 1(b). The PC in domain B possesses Dirac points at the K and K' points. Due to TRS breaking imposed by the external field, the Dirac points in domains A and C become gapped and carry opposite Chern numbers in their bandgap. Since domain B is not gapped, EM wave can propagate inside the bandgap (around 11.9 GHz) of domain A and C; but on its own, the propagation does not display any topological behavior. As domain B forms a domain wall between the two topologically distinct domains A and C, the principle of bulk-edge correspondence [17,18] implies that the combination of A, B and C domains can give rise to new topologically nontrivial signatures inside domain B with the number of unidirectional modes determined by the difference of gap Chern numbers in domains A and C. The projected band structure of the whole heterostructure is shown in Fig. 1(c). There are two (one is even and the other is odd about $y = 0$ plane) topological one-way large-area waveguide states (TOLWSs) in the bulk gap, which are uniformly extended in the whole domain B, as shown in the right panel of Fig. 1(c) by the intensity and phase distributions of the electric field. The one-way mode is distributed evenly over a large area, and hence the channel can carry a gigantic amount of energy without burning the device, in contrast to the edge modes in previous one-way structures [7-16] which are spatially tightly confined. As the waveguide is distributed over a wide area, we can conveniently use the structural degree of freedom to tune the information-carrying characteristics of the modes.

Fig. 2(a) shows the photo of the sample A|$B_9$|C. The subscript "9" indicates the width of the non-magnetized domain B, which has 9 layers in the *y*-direction. We can selectively excite the even or odd waveguide mode in domain B by polarized sources (electric dipole with polarization perpendicular to the *xy*-plane for the even mode, and magnetic dipole with polarization along *x*-direction for the odd mode). These two waveguide modes with different symmetries have the large separation in momentum space. The EM waves are injected into the sample at the point source labeled by a magenta star on the right boundary. The dispersion of each TOLWS can be obtained by measuring the field scanned along the *x*-direction at different frequencies and then performing a Fourier transform. The experimental results are plotted using color maps in Figs. 2(b) and 2(c) for the even and odd modes, respectively. Good agreements with the numerical results marked by the green lines are found.

It is remarkable that although domain B is a trivial crystal, it only allows unidirectional transmission in a certain frequency range. EM field can pass through the bulk of domain B when a point source (odd mode at 11.9 GHz) is placed on the right boundary, as shown in Fig. 3a. However, transmission is forbidden for the same A|$B_9$|C configuration when the point source is located on the left boundary as shown in Fig. 3(b). Strong transmission asymmetry is also observed at other passing frequencies of the TOLWSs, which is shown by the shadowed region in Fig. 3(d) where S21 and S12 denote the measured transmission spectra averaged along the *y*-direction for the two cases. Strong contrast between the two transmissions is clearly seen inside the passing frequency window of TOLWSs. Figure 3(e) shows the transmitted amplitude measured point-by-point along the white dotted lines 1 and 2 in Fig. 3(a) and (b) for these two cases. These results provide strong evidence of one-way transport in such heterostructures.

The TOLWSs here has a much larger and tunable transmission area than an ordinary topological one-way edge mode [7-16], which is exponentially confined to the boundary. The TOLWSs are also very robust against defects. Taking advantage of these characteristics, we can achieve topological one-way beam spatial collimation using the structure shown in Fig. 3(c), in which the layer number in domain B is sharply reduced from 9 to 0 (from right side to left side). If there were no topological protection, the waves would have been strongly reflected when the transport channel is abruptly narrowed. However, the absence of backscattering channels (as evident from Fig. 1(c)) forces the waves to go forward and all the energy is squeezed from the wide channel into the narrow channel. The simulation result of topological one-way beam spatial collimation effect is shown in Fig. 3(c), where an extended TOLWS excited by a point source on the right is collimated into a narrow channel on the left due to the sudden narrowing of the transport channel. Fig. 3(f) compares the intensity of the waves along the white dot lines 3 (Fig. 3(c)) and 1 (Fig. 3(a)), showing that the intensity in the narrow channel is increased several times. The wide and tunable width of TOLWSs allows for a high efficiency collection of energy from an extended region and guides all the captured energy into a narrow channel with high intensity.

In Figs. 4 and 5, we further show the robustness of TOLWS transport against various kinds of defect and randomness. We create a void in the middle of the B domain by removing 10 unit cells as marked by the green dashed rectangle in Fig. 4(a) and a PEC obstacle in Fig. 4(b). A point source (exciting the odd mode) marked by the magenta star located on the right side of domain B is used to excite the TOLWS. In Fig. 4(a) and Fig. 4(b), the simulation results of electric field intensity distribution at 11.9 GHz show almost complete recovery of the TOLWS after passing through the area with the void or PEC obstacle, which appear almost identical to that shown in Fig. 3(a) without defects, indicating that the void or PEC obstacle is almost invisible to the TOLWS. Fig. 4(c) and Fig. 4(d) show the measured spectra of average transmission S21 and S12 obtained by placing the point source on the right and left boundaries of the domain B for these two cases, respectively. Remarkably, inside the passing frequency window of the TOLWSs, the spectra are also found to be nearly identical to those shown in Fig. 3(d) without defects. It should be stressed that the robustness of the one-way transport of the TOLWSs is a result of time-reversal symmetry breaking in domains A and C, with the consequence that both the TOLWSs transmit energy in the same direction (see the green lines in Fig. 1(c)) and hence no backscattering channels exist. This is very different from the valley transport produced by parity breaking [19,20,22-25], in which the transport can be greatly compromised by the presence of defects due to the presence of a backscattering channel.

Now we study the transport of TOWLSs in the presence of randomness in domain B. It is well known that all EM waves will be localized in random two-dimensional systems due to Anderson localization [26-30]. The localization effect is demonstrated in the four upper panels of Fig. 5(a) at the frequency of 11.7 GHz. The first panel shows the perfect transmission for an ordered PC. The other three panels show the complete blocking of waves by the three different random configurations of the YIG rods inside the PC, with the incident wave coming from the right hand side of the channel. Here, randomness is introduced inside the green rectangle by randomly assigning the permittivity of YIG rods between 1 and 26.6. However, if the PC B is sandwiched by domain A and C as shown in Fig. 1(a), the transport of TOWLSs is protected. It cannot be hindered by the presence of defects or randomness due to the absence of backscattering channel and the accompanying coherent backscattering effect. The simulation results of the transport of TOWLS at 11.7 GHz are shown in the second row panels of Fig. 5(a). In the case of no randomness, the first panel shows again the perfect transmission with uniform field distribution inside the B domain. When randomness is introduced, the other three panels show the irregular field distributions inside the B domain. However, the transport of waves is largely protected with an average transmission close to unity. Similar results are shown for the case of 11.9 GHz. In Fig. 5(b), we plot the average transmission spectra in log scale for both the ordered case and the three random cases. Close to unity transmission is found in the heterostructure ABC systems for all three random cases.

The robustness of the transport of TOLWSs against defects or randomness in domain B can be understood from the formation of TOLWSs. It is known that the application of magnetic field in domain A and C will introduce one-way interface states at the interfaces of AB PCs and BC PCs. Since the applied fields in domains A and C are in opposite directions, the two one-way interface states now propagate in the same direction. The presence of the B domain reduces the gap size of the PCs A and C. As domain B itself does not have a gap, the two topological interface states (one from A and one from C) do not decay inside domain B and they couple strongly and hybridize into a pair of even/odd waveguide states with almost uniform field amplitude inside the waveguide. Thus, the pair of waveguide states owe their topological properties to the Chern numbers in the domains A and C, even though domain B is topologically trivial. It can be viewed as a form of proximity effect, but oddly, domain B (the width of the waveguide) can be as big as one likes, at the expense of a smaller operational bandwidth for one-way transport. The presence of defects or randomness in domain B cannot hinder the wave transport of TOLWSs. We note again that the merging of two interface states into waveguide states does not depend on the width of the B domain. As the width of the B domain is increased, the size of the gap in the heterostructure will also reduce, which in turn will increase the decay length of the interface states and keeps the two interface states connected.

In conclusion, we have realized and experimentally observed the TOLWSs by designing heterostructures using magnetic PCs. The TOLWSs cover a large area and hence have more degrees of freedom to the tune energy transport capacity and characteristics. As a result, unique topological one-way beam spatial collimation effect is demonstrated. As TRS is broken in this system, the ABC structure is robust to different kinds of defect and randomness. The heterostructures proposed here provide a pair of robust one-way waveguide modes for wave transport. As the waveguide can be as wide as we like, it also provides an efficient way to collect information and energy.

This work was supported by the Hong Kong Research Grants Council under Grant No. AoE/P-02/12 and 16303119. Lei Zhang was supported by National Natural Science Foundation of China Grants No. 11704232, National Key R\&D Program of China under Grants No. 2017YFA0304203, 1331KSC, Shanxi Province 100-Plan Talent Program. We thank Prof. Xu-Lin Zhang for helpful discussions.

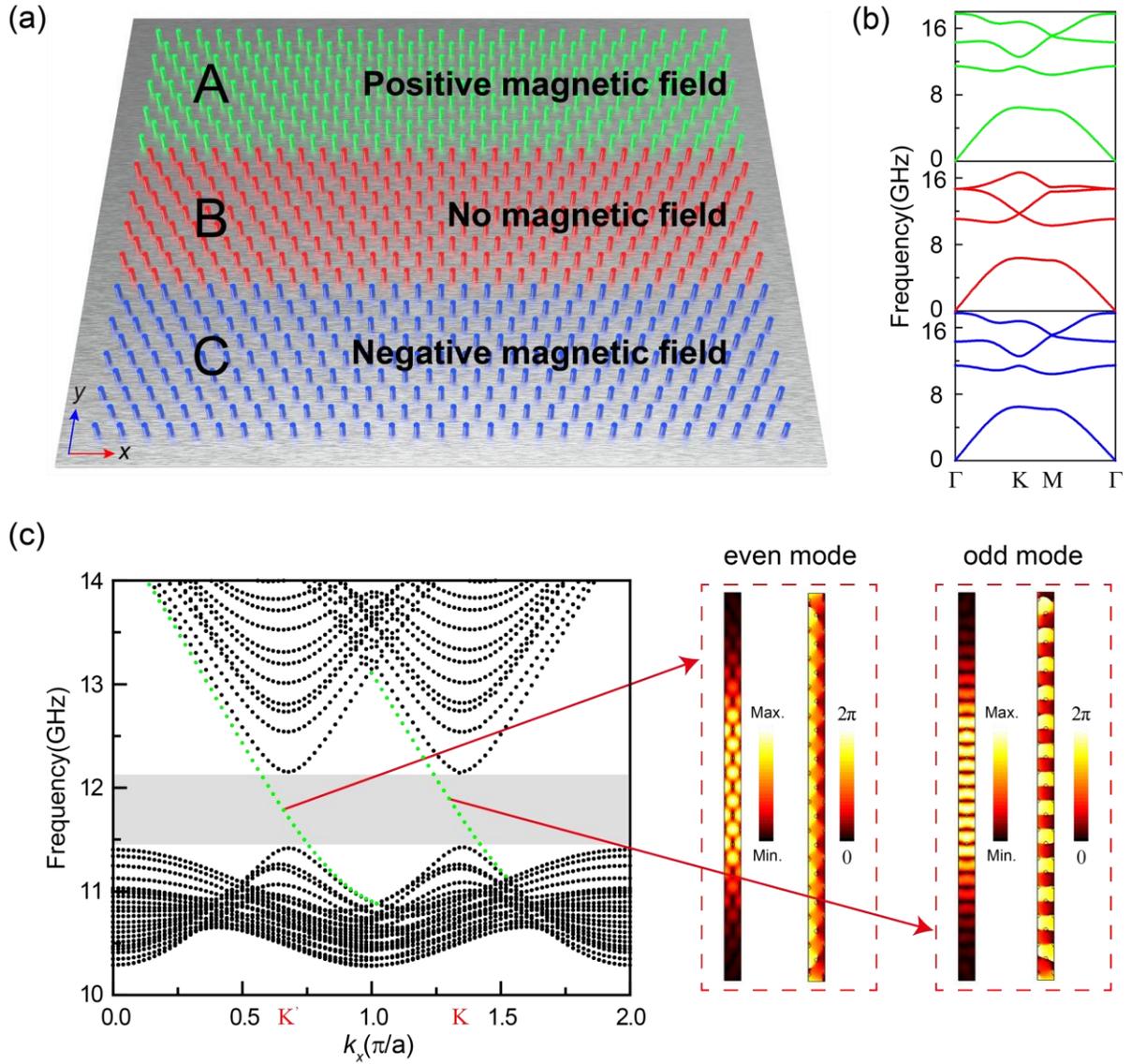

Fig. 1. (a) The schematic picture of the A|B$_9$|C heterostructure. Positive (negative) magnetic field is added in Domain A (C), and no magnetic field is added in Domain B. (b) Bulk band structures for A, B and C, respectively. (c) Left panel: the projected band structure for A|B$_9$|C. The shadowed area marks the frequency range of the TOLWSs, which are indicated by two green dotted lines. Right panel: the intensity and phase distributions of electric fields of two TOLWSs in domain B, respectively.

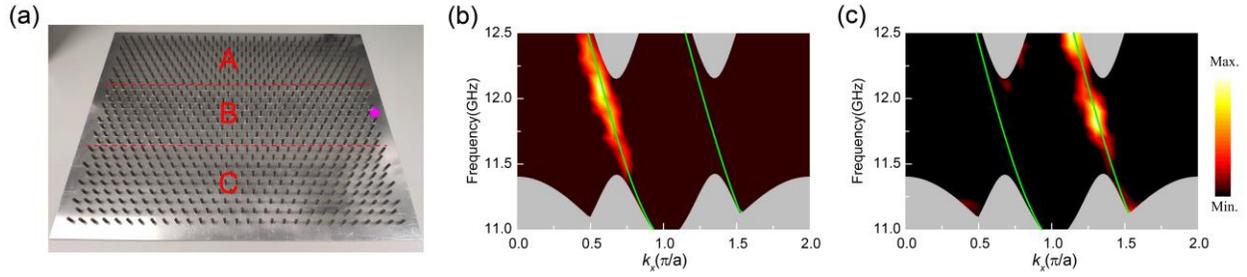

Fig. 2. (a) The photo of the A|B$_9$|C sample. The magenta star marks the position of the point source. (b),(c) The selective excitations of the even (b) and odd (c) TOLWSs by the corresponding point dipolar sources.

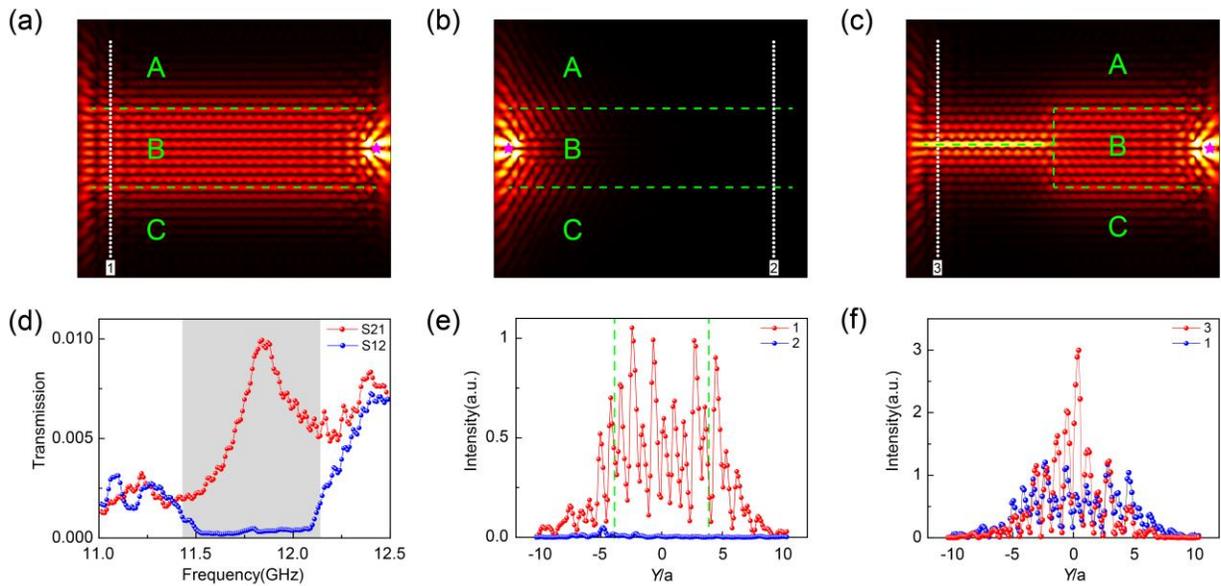

Fig. 3. (a),(b) Simulated electric field intensity distributions of excited by point sources (magenta stars) on the right and left sides for the same A|B$_9$|C configurations at 11.9 GHz, respectively. (c) Simulated electric field intensity distribution using a point source excitation at 11.9 GHz on the right boundary of the B domain. Its width is sharply reduced from 9 layers to 0. (d) S21 (S12) is the measured transmission spectrum averaged along the transverse direction in the B domain when the point source is placed on the right (left) boundary. (e) Red (blue) solid line is the experimental intensity profiles at 11.9 GHz along the white dotted line 1 (2) in Fig. 3(a) (Fig. 3(b)). The area between two green dashed lines represents the domain B. (f) The red (blue) solid line shows the experimentally measured intensity profiles along the white dotted line 3 (1) in Fig. 3(c) (Fig. 3(a)).

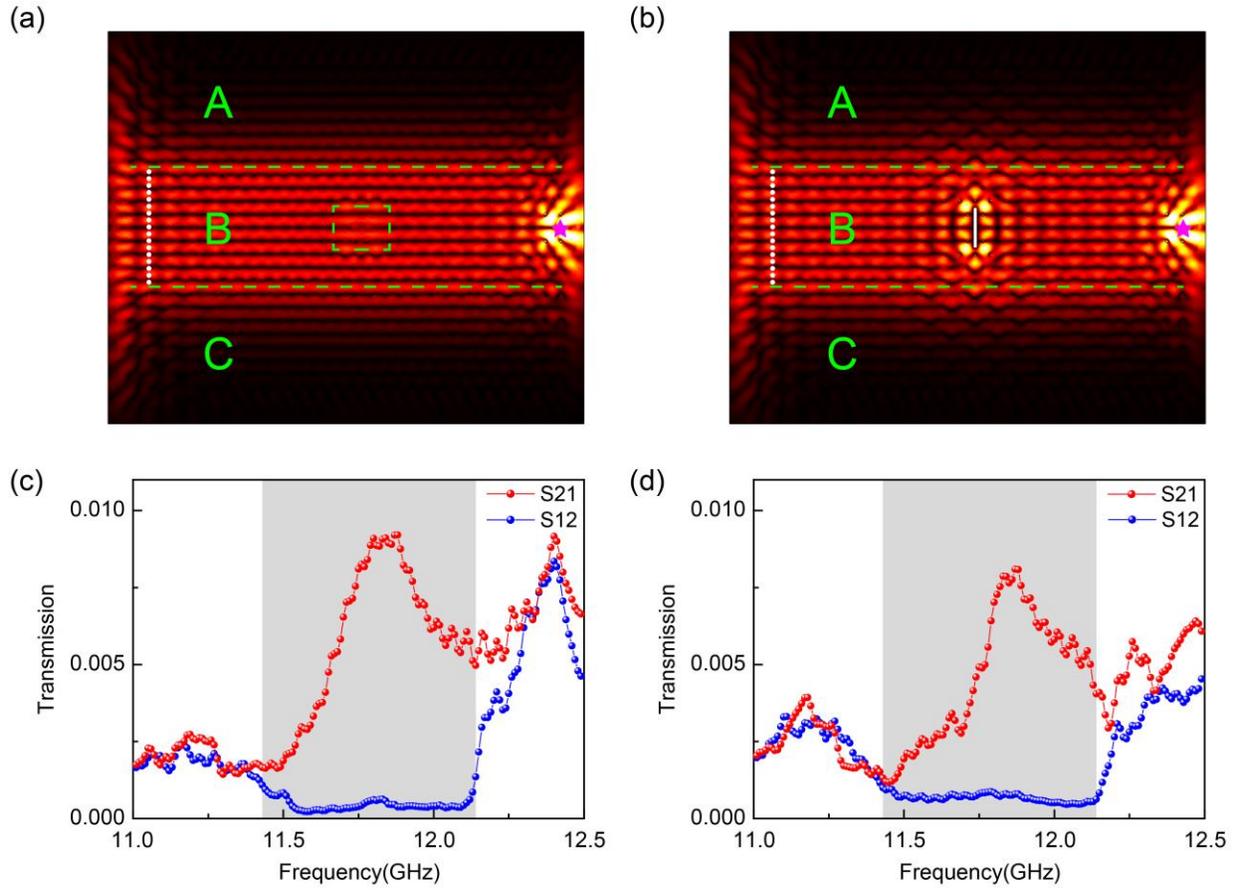

Fig. 4. (a),(b) Simulated electric field intensity distributions at 11.9 GHz in the heterostructures with a void defect, in which 10 unit cells are removed inside the green dashed rectangle and a PEC obstacle, respectively. (c),(d) Measured average transmission spectra S21 and S12 in domain B for the void and PEC defects, respectively.

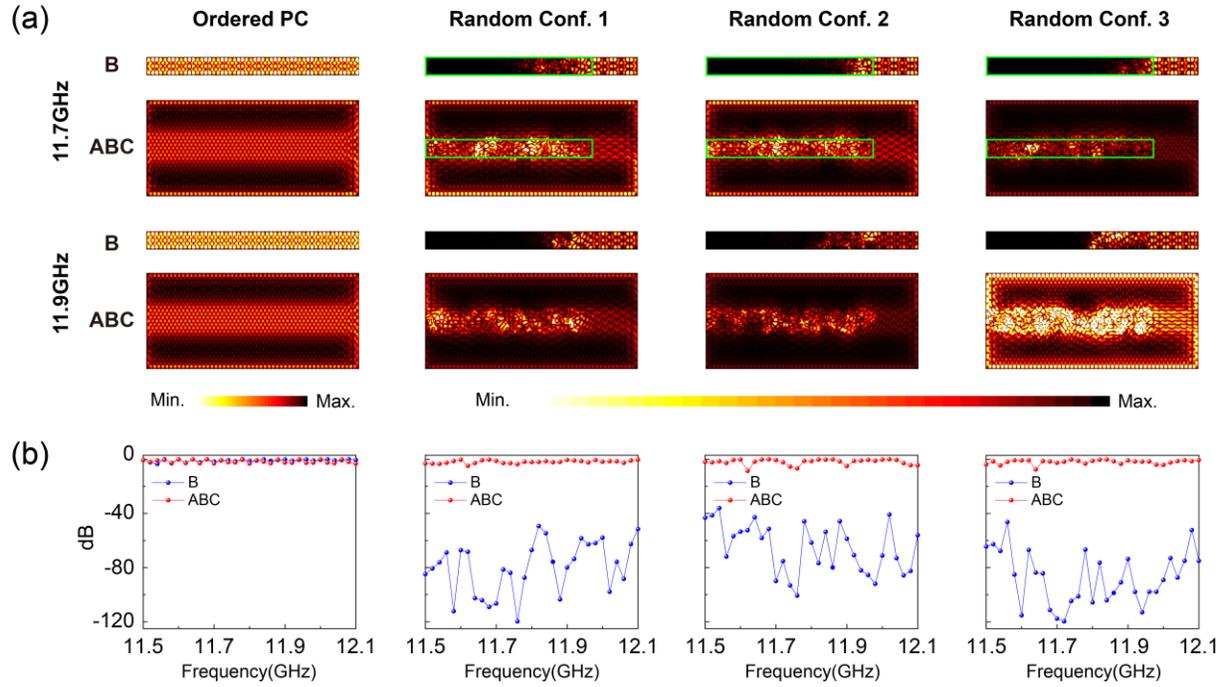

FIG. 5. (a) The simulations of a crystal $B_5$ and heterostructures $A|B_5|C$ in the ordered PC and three different configurations of random PCs, in which the permittivity of YIG rods inside the green rectangle is randomly selected between 1 and 26.6. Electromagnetic waves are incident from the right. (b) The corresponding transmission spectra for $B_5$ and heterostructures $A|B_5|C$.